\documentclass[useAMS,usenatbib]{mn2e}
\usepackage{graphicx}
\usepackage{rotating}
\usepackage{times} 
\title[Molecular content of a type-Ia SN host galaxy at
$z=0.6$]{Molecular content of a type-Ia SN host galaxy at
$z=0.6$}
\author[A.-L. Melchior $\&$ F. Combes ]{A.-L. Melchior,$^{1,2}$\thanks{E-mail:
A.L.Melchior@obspm.fr} F. Combes$^{1}$ \\ $^{1}$LERMA, Observatoire de Paris,
UMR8112, 61, avenue de l'Observatoire, Paris, F-75014, France\\$^{2}$ Universit\'e Pierre et Marie Curie-Paris 6, 4, place Jussieu, F-75252 Paris Cedex 05, France\\}
\begin{document}
\date{Accepted ; Received }

\pagerange{\pageref{firstpage}--\pageref{lastpage}} \pubyear{2007}

\maketitle

\label{firstpage} 

\begin{abstract}
We study the properties and the molecular content of the host of a
type-Ia supernova (SN1997ey). This $z=0.575$ host is the brightest
submillimetre source of the sample of type-Ia supernova hosts observed
at 450$\mu$m and 850$\mu$m by \citeauthor{Farrah:2004}. Observations
were performed at IRAM-30m to search for CO(2-1) and CO(3-2) lines in
good weather conditions but no signal was detected. The star formation
rate cannot exceed 50 $M_\odot$/yr. These negative results are
confronted with an optical analysis of a Keck spectrum and other data
archives.  We reach the conclusion that this galaxy is a late-type
system (0.7~$L^B_*$), with a small residual star-formation activity
(0.2~M$_\odot$\,yr$^{-1}$) detected in the optical. No source of
heating (AGN or starburst) is found to explain the
submillimetre-continuum flux and the non-CO detection excludes the
presence of a large amount of cold gas. We thus suggest that either
the star formation activity is hidden in the nucleus (with $A_V\sim
4$) or this galaxy is passive or anemic and this flux might be
associated with a background galaxy.
\end{abstract}


\section{Introduction}
Standardised type-Ia supernovae (SN) have been extensively studied to
probe the expansion of the Universe for the past decade (e.g. Riess,
Press $\&$ Kirshner \citeyear{Riess:1996};
\citealt{Perlmutter:1997,Perlmutter:1999,Astier:2006}). The mechanisms
that rule these cosmic explosions are not fully understood. A better
knowledge of their environment and of their host galaxies is important
to understand possible systematics, which could affect these
cosmological probes \citep[e.g.][]{Combes:2004}, and the progenitor
systems, which are also important for the evolution of galaxies
\citep[e.g.][]{Hamuy:2003,Panagia:2006,Howell:2007}.  The known
scatter of the SNIa standard candles is corrected empirically on the
basis of the observed tight correlation between the peak luminosity
and the decline rate of the light curve
\citep{Phillips:1993,Riess:1996,Perlmutter:1997}. Standardised SNIa
detected at high-z do not exhibit any sign of residual extinction
\citep[e.g.][]{Riess:1998,Perlmutter:1999,Farrah:2002,Sullivan:2003},
as selection effects most probably eliminate the most obscured SN.

A submillimetre survey of 31 SN-Ia hosts
\citep{Farrah:2004,Clements:2005} has detected two strong sources at
850~$\mu$m at the $7\sigma$ level. It was surprising to find
submillimetre-bright galaxies in this sample of SN-Ia
hosts. Nevertheless, this strengthens the observation of the
correlation of the SNIa rate with the Star-Formation Rate (SFR)
\citep[e.g.][]{Sadat:1998,Sullivan:2006}, and the evidence 
(e.g. Mannucci, Della Valle $\&$ Panagia \citeyear{Mannucci:2006};
\citealt{Sullivan:2006}) of the possible association of one type of
SNIa with recent star-formation. In addition, according to their
optical morphology \citep{Farrah:2004}, these submillimetre-bright
hosts look like ordinary disc galaxies. In order to try to better
understand the nature of these hosts, we try to observe the CO lines
of SN1997ey host, whose continuum has also been detected at 450~$\mu$m
at the $6\sigma$ level. 
\begin{figure*}
\includegraphics[width=0.55\textwidth,angle=-90]{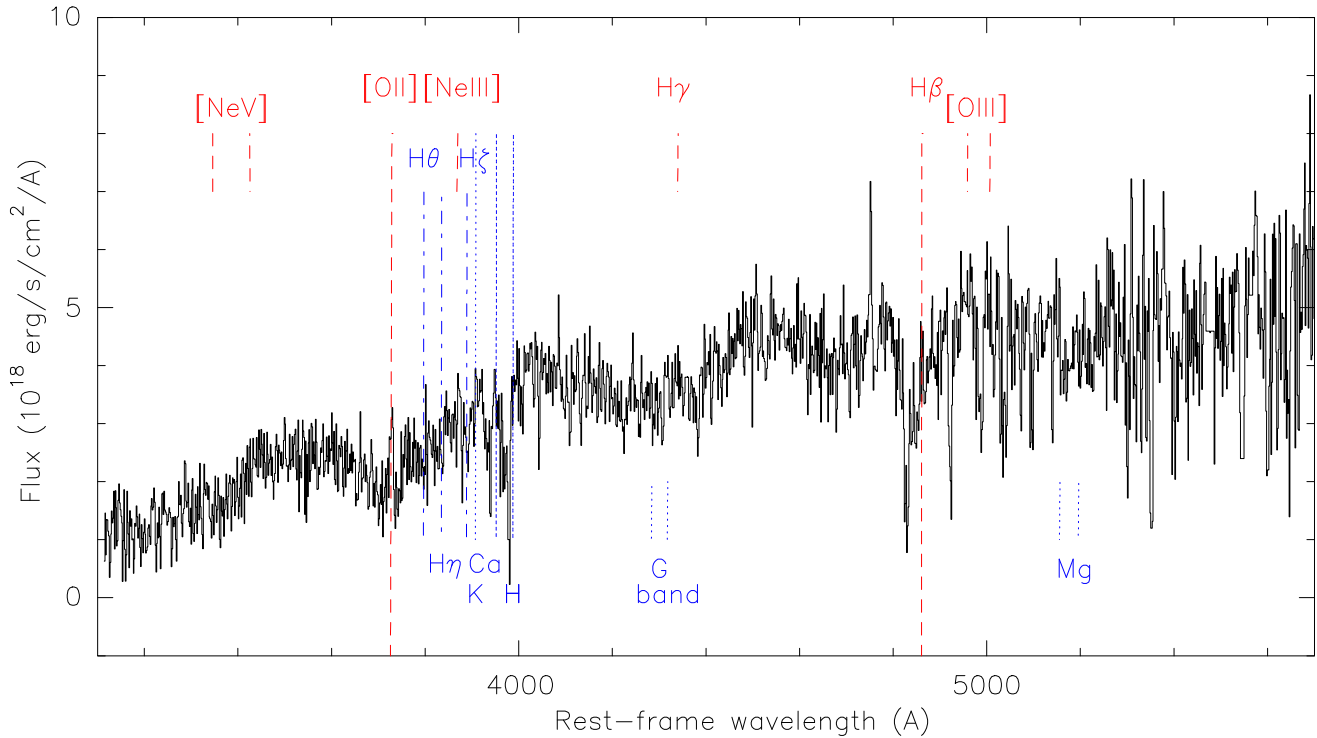}\\
\caption{Optical spectrum obtained by the Supernova Cosmological
Project (2007, private communication) on the Keck telescope on 1997
December 31 (06:22 UT), while the supernova SN1997ey has been detected
on 1997 December 29 \protect\citep{Nugent:1998}. This spectrum
positioned on the supernova, containing both the host galaxy ($R\sim
21.7$) and the supernova fluxes ($R\sim 22.9$), provides a secure
redshift at \protect$z=0.575$. The detected (resp. marginal or close
to detection) emission lines are indicated with long (resp. short)
(red) dash lines ([NeV], [OII], [NeIII], H$\gamma$, H$\beta$, [OIII])
while absorption features are displayed with long (resp. short) (blue)
dotted lines (Ca H and K, G band, Mg) or dash-dotted lines (H$\zeta$,
H$\eta$, H$\theta$)
\citep[see e.g.][]{McQuade:1995}. The brightest night sky lines
\citep{Osterbrock:1996} have been removed.}
\label{fig:spectra}   
\end{figure*}

In section \ref{label:charac}, we discuss the characteristics of this
host galaxy relying on data archives. In section \ref{label:CO}, we
present the CO observations performed at IRAM-30m. In section
\ref{label:discussion}, we discuss these results.

Throughout this paper, we adopt a flat cosmology, with
$\Omega_\rmn{m}=0.24$, $\Omega_\Lambda=0.76$ and
$H_0=73$~km\,s$^{-1}$\,Mpc$^{-1}$ \citep{Spergel:2007}.

\section[]{Characteristics of SN1997ey host}
\label{label:charac}
SN1997ey host has been initially detected as the host of a type Ia
supernova \citep[SN1997ey,][]{Pain:2002}. It was first detected with
ground-based photometry ($R=21.7$, Pain 2004, private
communication). Its spectroscopic redshift ($z=0.575$) was determined
with the Keck telescope \citep{Nugent:1998}. It was then observed by
{\it HST}/STIS ({\it HST} Proposal 8313, Ellis (1999), see also
Fig. \ref{fig:imaging}). More recently, investigating its dust
content, \citet{Farrah:2004} detected the submillimetric continuum of
this galaxy at 450 and 850~$\mu$m.  We review in this section these
properties in more details.

\subsection[]{Optical spectroscopy and detection of SN1997ey}
\label{ssec:opticalspec}
\begin{figure}
\includegraphics[height=0.48\textwidth,angle=270]{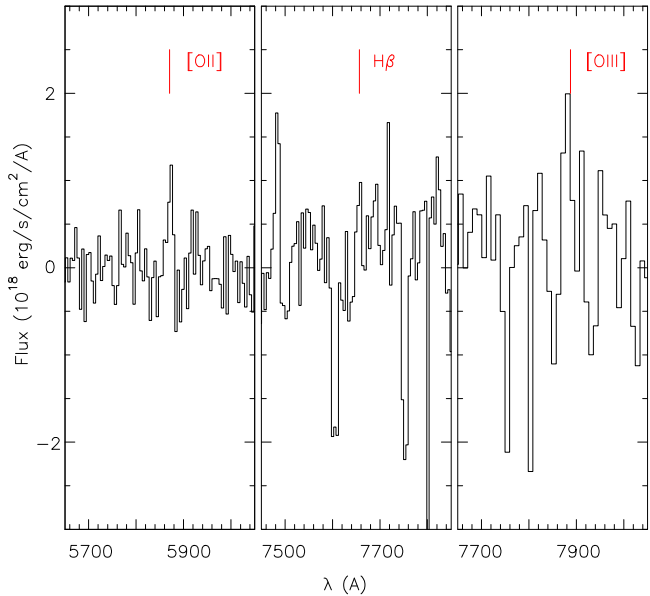}
\caption{Emission lines used to estimate the $[$O\,II$]$ (3727~$\AA$),
H$_\beta$ (4861~$\AA$) and $[$O\,III$]$ (5007~$\AA$)
luminosities. The measured flux, after baseline subtraction, is
displayed in function of the observed wavelength. From these
luminosities, we derived estimates of the SFR and of the metallicity
of this host galaxy (see Section \ref{ssec:opticalspec}).  }
\label{fig:spectralines}   
\end{figure}
The Supernova Cosmology Project (2007, private communication) took a
spectrum of the supernova close to maximum with Low Resolution Imaging
Spectrometer (LRIS) on the Keck telescope on 1997 December 31, as
displayed in Fig. \ref{fig:spectra}. The magnitude of the supernova at
discovery is $R=22.9$, while the $R$ magnitude of the host is $R\sim
21.7$.  This spectrum, obtained with the 1\arcsec-wide slit aligned on
the galaxy centre and the SN position, is a combination of the SN
($\sim 1/3$) and the host galaxy ($\sim 2/3$) spectra (SCP, Hook 2007,
private communication). The host spectrum contains stellar absorption
lines and some emission lines (see Fig.
\ref{fig:spectra}).  In Fig. \ref{fig:spectralines}, we focus on
$[$O\,II$]$, H$\beta$ and $[$O\,III$]$ emission lines to characterise
the properties of the host. We expect that most of the galaxy flux is
contained in this spectrum given the position of the slit. Moreover,
most of the star-formation activity usually lies in the central part
of the galaxy or at least in the galactic plane. Last, if we assume
that this spectrum continuum contains only 2/3 flux from the host and
scale the spectral continuum to match the observed broad band flux in
the $R$ band (factor $1.7$), we should multiply the whole spectrum by a
factor 2/3$\bmath{\cdot} 1.7=1.1$. In the following, we estimate that
the extra 10 per cent lies within the uncertainties of this procedure
and work directly on the spectrum displayed in Fig.
\ref{fig:spectra}.
We find $L([\rmn{O\,II}])=2.5\bmath{\cdot} 10^{33}$~W,
$L(\rmn{H}\beta)=2.9\bmath{\cdot} 10^{33}$~W and
$L([\rmn{O\,III}])=7.8\bmath{\cdot} 10^{33}$~W. Relying on Kewley,
Geller $\&$ Jansen \citeyearpar{Kewley:2004}, this indicates a
moderate on-going star-formation activity of 0.2~M$_\odot$\,yr$^{-1}$
(with no extinction correction) and a solar metallicity. The other
Balmer lines are not detected in emission (but possibly in
absorption), which might suggest some extinction.  In parallel, the
Ca\,II H and K absorption lines are clearly detected and the ratio of
Ca\,II\,H and H$\epsilon$ to Ca\,II K is larger than unity, typical
for stars with spectral type later than F \citep{Rose:1985}.  This
host contains a significant population of stars older than $\sim
1$~Gyr \citep[e.g.][]{Delgado:2005}.

We derived the rest-frame B luminosity from the R
magnitude of the host $L_B^{\mathrm{host}} \sim 0.15 
\bmath{\cdot} 10^{10} \rmn{L}_\odot \sim 0.7 \bmath{\cdot} L^B_*$,
where $L_*$ is defined with the Schechter function
\citep[$L^B_*=2.1\bmath{\cdot} 10^{9}\rmn{L}_\odot$ in][]{Marzke:1998}. 
Unfortunately, this supernova was not monitored after detection, so no
light curve is available. The supernova type and phase were determined
by fitting the spectrum with SN and galaxy spectral templates, as
described in \citet{Howell:2005}.  The best matches were all SNe Ia
with a mean epoch of +2 days and a scatter of 6 days.  SN1997ey is
located 2.77\,\arcsec\, from the centre of the host galaxy, which
corresponds to a projected distance of 18\,kpc \citep[see fig. 1
of][]{Farrah:2004}. Only 2 out of a sample of 15 host galaxies studied
by \citet{Farrah:2002} are detected at an offset larger than 15~kpc
and both occurred in E/S0 host galaxies.

GRB971221 has been detected at RA 73.7$\degr$ and DEC 4.7$\degr$, with
an error box (BATSE) of 6.3$\degr$. The association of SN1997ey and
GRB971221, suggested by \citet{Bosnjak:2006}, relies on this
inaccurate position but also on the time coincidence of the 2
events. As discussed in section \ref{sec:grb}, this association is
most probably a chance alignment.

\subsection[]{Imaging}
\begin{figure}
\centering
\includegraphics[width=0.48\linewidth]{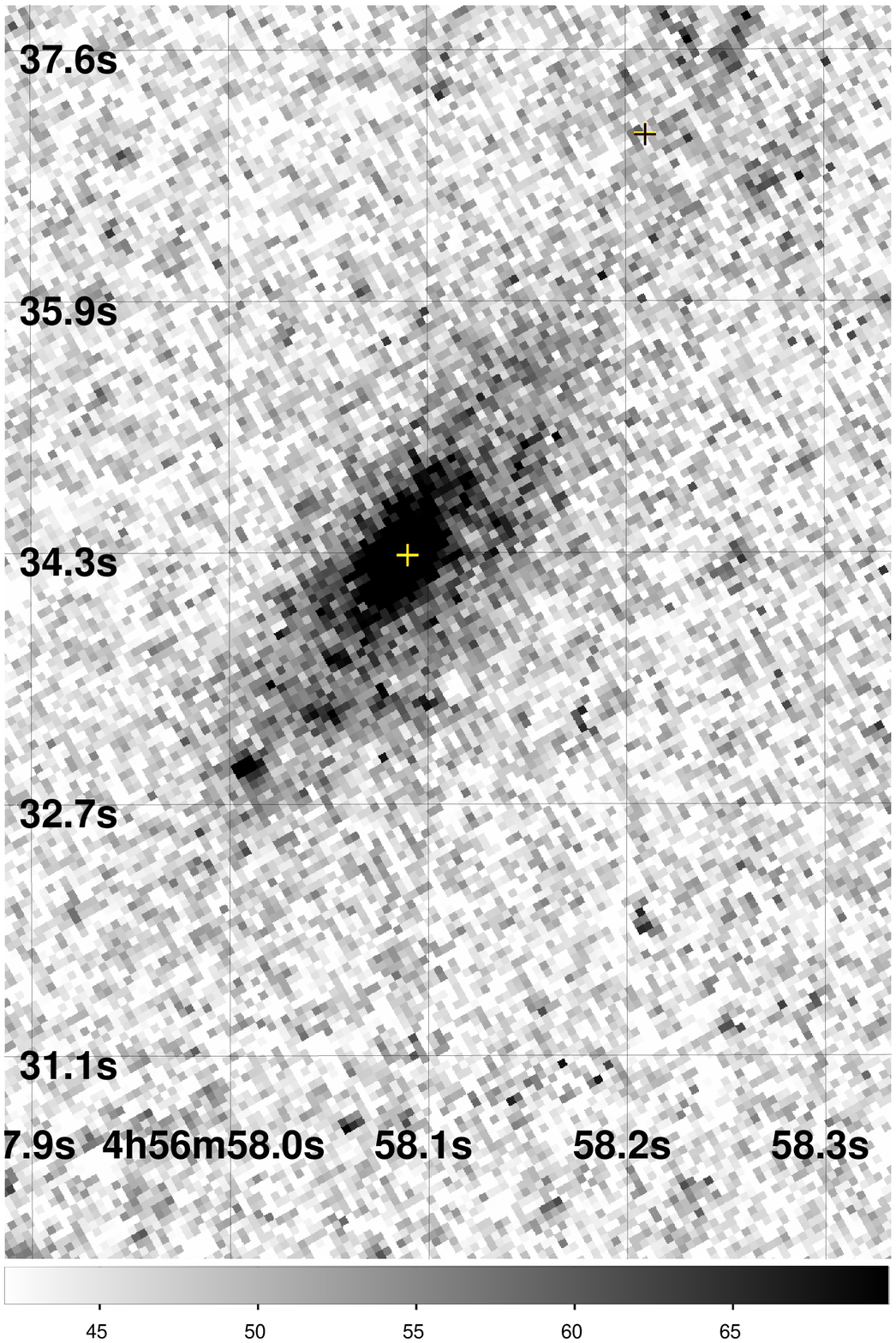}
\includegraphics[width=0.48\linewidth]{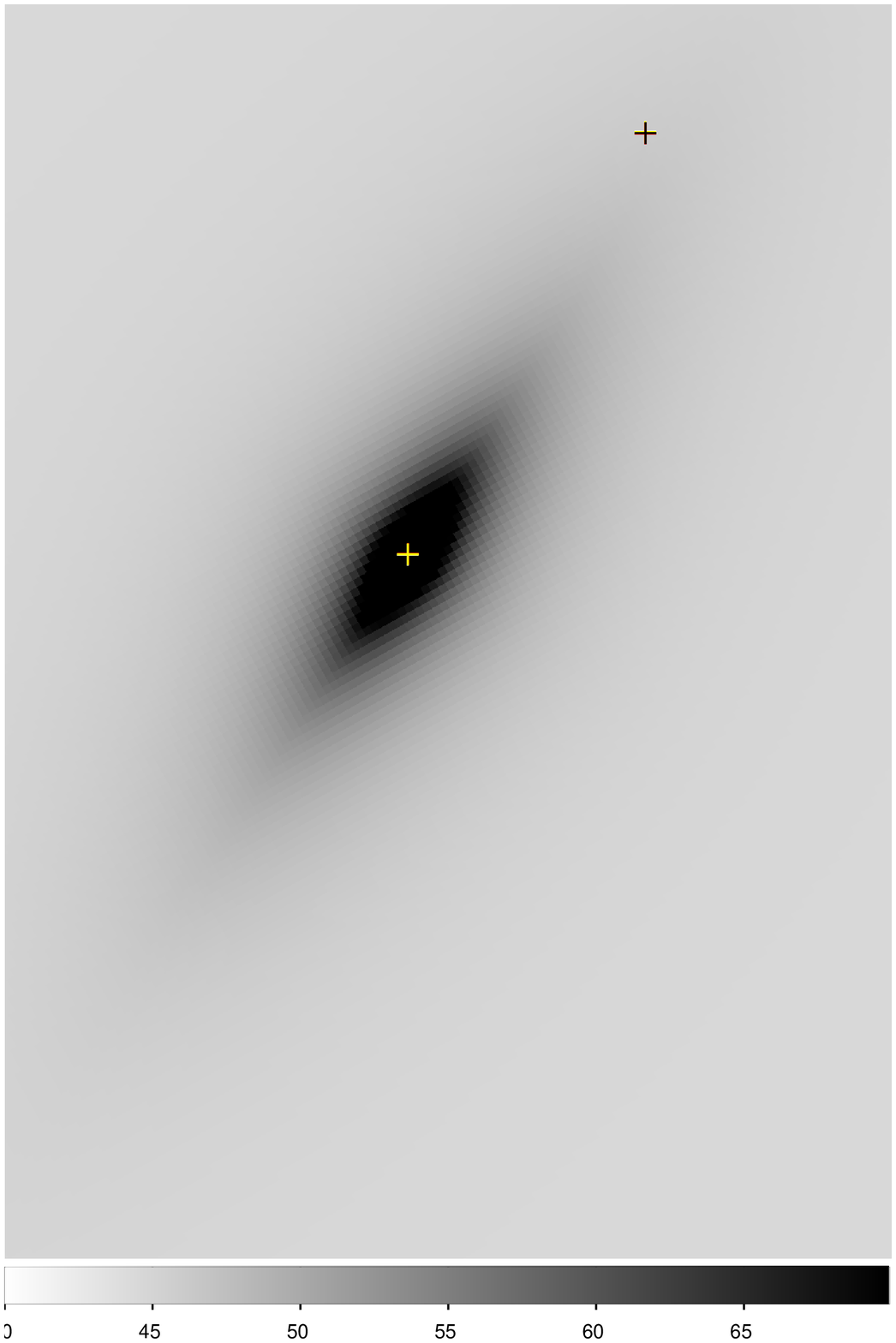}
\caption{HST/STIS image of SN1997ey host. The left panel displays the
optical image of this galaxy. The right panel shows the modelling
obtained by {\sc GALFIT}.  On each panel, the crosses indicate
respectively the positions of the galaxy centre (J2000
04$^h$56$^m$58.1$^s$ -02$\degr$ 37$\arcmin$34.3$\arcsec$) and of SN1997ey
(J2000 04$^h$56$^m$58.2$^s$ -02$\degr$37$\arcmin$37$\arcsec$).}
\label{fig:imaging}   
\end{figure}
We retrieve from the HST archive the STIS image obtained by R. Ellis
in 1999 (Proposal, 4647). We reduce the three available exposures with
IRAF and used the drizzle procedure to remove the cosmic rays
\citep{Fruchter:2002}. In order to better understand the nature of
this host, we try to fit different profiles to the STIS/HST image
using the software package {\sc GALFIT} \citep{Peng:2002}.  The best
fit was obtained for one S{\'e}rsic profile with an effective
(half-light) radius $r_e=1.8\arcsec=11$~kpc, a power-law index
$n=2.04$, a diskiness parameter $c=-0.45$ and an axis ratio
$b/a=0.39$. We thus estimate an inclination of order 70$\degr$
\citep{Paturel:1997}. The 10-20 per cent residuals are asymmetric
along the major axis. This disc structure suggests an inclined Sc
spiral galaxy.  There are several galaxies ($\sim 20$) in the field of
view ($45\arcsec$ per $45\arcsec$) of this image, so this host might
be member of a group or a cluster.

\subsection[]{Optical-to-radio continuum spectrum}
\label{ssec:opticaltoradio}
\begin{figure}
\centering
\includegraphics[width=8cm]{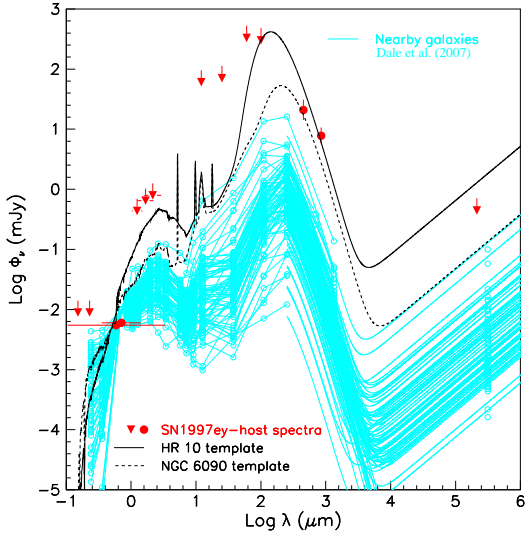}\\
\caption{Host galaxy continuum spectrum of the type-Ia SN SN1997ey. The
(red) symbols correspond to direct detections (bullets) and upper
limits (arrow) obtained from various archives. The (red) data points
display the detection and upper limits derived {\it GALEX}, SCP, {\it
HST}/STIS, 2MASS, {\it IRAS}, SCUBA \citep{Farrah:2004} and NVSS. See
details in the text. The (light blue) connected data points are from
\protect\citet{Dale:2007}. The full-line and dashed-line templates
correspond to fits to the starburst galaxies: HR10
\protect\citep{Dey:1999} and NGC6090
\protect\citep{Calzetti:2000}. They correspond respectively to a star
formation rate of 373\,M$_\odot$\, yr$^{-1}$ and 32\,M$_\odot$\,
yr$^{-1}$.}  
\label{fig:sn1997eyhost}   
\end{figure}

We derive the continuum spectrum from data archives (see Table
\ref{tab:data}) as displayed in Fig. \ref{fig:sn1997eyhost}. We
retrieve data points in the optical and in submillimetre wavelengths
as follows: (1) In the optical, one point was obtained from the SCP
$R=21.7$ measurement and the other was derived from a compilation of 3
'unfiltered' images ($\langle \lambda\rangle = 5861.5\AA$ with
FWHM$=4410~\AA$) from the {\it HST}/STIS archive. They are both
consistent. (2) In the submillimetre range, we use the $7\sigma$ and
$6\sigma$ SCUBA detection \citep{Farrah:2004} at 850 $\mu$m and
450$\mu$m. We then derive upper limits from the released all-sky
surveys, which observed this position, namely {\it GALEX} at 150\,nm
and 227.5\,nm, 2MASS in $J$, $H$ and $K_s$ \citep{Skrutskie:2006},
{\it IRAS} (10, 25, 60 and 100~$\mu$m) and NVSS (NRAO VLA Sky Survey)
at 1.4~GHz
\citep{Condon:1998}.

We then compare this continuum spectrum with various templates. First,
we superimpose the measurements of the nearby galaxy sample of
\citet{Dale:2007}, that we normalise to $0.01$~mJy at $4500~\AA$
and shift to the appropriate redshift.  Second, we add starburst
templates obtained by \citet{Melchior:2001} by fits to HR10
\citep{Dey:1999} and NGC6090 \citep{Calzetti:2000}, respectively blue
and redshifted to $z=0.575$.
\begin{table}
 \centering
\begin{minipage}{85mm}
  \caption{\protect Data points retrieved from various archives. We
provide the observed fluxes ($\Phi_\nu$) and associated error
($\Delta\Phi_\nu$) or the $3\sigma$ upper limits obtained or derived
from the mentioned archives.}
\begin{tabular}{llllll} 
\hline
$\lambda$  & $\Phi_\nu$  &  $\Delta\Phi_\nu$  & $3\sigma$
upper limits & Archives\\
($\mu$m) & (mJy) & (mJy) & (mJy)&\\
\hline
0.15             &                     &      & 8.9 $\bmath{\cdot} 10^{-3}$
&  {\it GALEX}\\
0.23             &                     &      & 8.9 $\bmath{\cdot} 10^{-3}$
&  {\it GALEX}\\
0.59             & 5.5$\bmath{\cdot} 10^{-3}$ &      &
&   {\it HST}/STIS\\
0.69             & 6.0$\bmath{\cdot} 10^{-3}$ &      &                       &  SCP\\
1.2              &                     &      & 0.44                  & 2MASS\\
1.7              &                     &      & 0.65                  & 2MASS\\
2.2              &                     &      & 0.79                  & 2MASS\\
12               &                     &      & 60.                   & {\it IRAS}\\
25               &                     &      & 71.                   & {\it IRAS}\\
60               &                     &      & 331.                  & {\it IRAS}\\
100              &                     &      & 302.                  & {\it IRAS}\\
450              &               20.80 & 3.54 &                       & SCUBA (*)\\
850              &                7.80 & 1.10 &                       & SCUBA (*)\\
214$\bmath{\cdot} 10^3$ &                     &      & 0.45                  & NVSS\\
\hline
\end{tabular}
\end{minipage}
(*) Farrah et al. (2004)
\label{tab:data}
\end{table}
The templates correspond to $L_{\rmn{IR}}=1.84\bmath{\cdot}
10^{11}~\rmn{L}_\odot$ for NGC6090 and $L_\rmn{IR}=2.17\bmath{\cdot}
10^{12}~\rmn{L}_\odot$ for HR10, corresponding to star-formation rates
(SFR) of 32~M$_\odot$\,yr$^{-1}$ and 373~M$_\odot$\,yr$^{-1}$
\citep{Kewley:2002}.

\vspace{1cm}

This submillimetre-bright galaxy with a secure spectroscopic redshift
($z=0.575$) is a good candidate to search for CO. As shown by
\citet[see their fig. 3 ]{Greve:2005}, optical spectroscopic
redshifts are usually in good agreement with the CO-determined
redshift within 75~km\,s$^{-1}$.  While few CO investigations are
conducted so far in this redshift range \citep{Greve:2005} due to the
lack of CO lines in the 3\,mm window, we searched for CO(2-1) and
CO(3-2) lines in this galaxy thanks to the 2-mm and 1-mm receivers
available at the IRAM-30m telescope.

\section[]{CO observations and reduction of the data}
\label{label:CO}
\begin{figure}
\centering
\includegraphics[width=8cm]{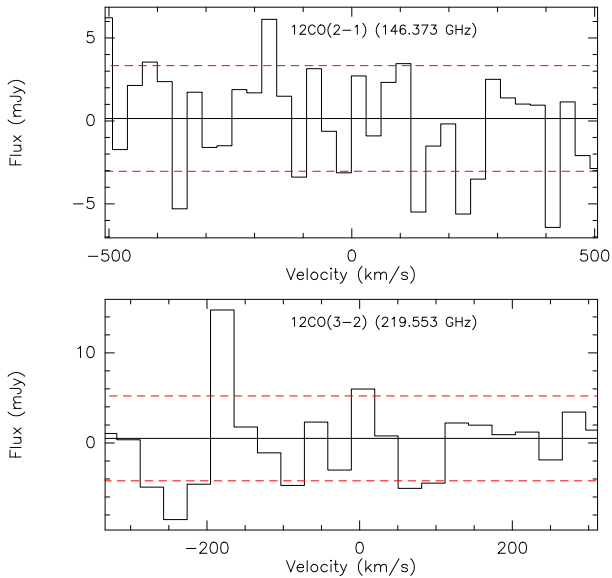}\\
\caption{Non-detection of the CO(2-1) and CO(3-2) lines searched at
IRAM-30m (2004 July 7-11) at the spectroscopic redshift ($z=0.575$) of the host
galaxy. The displayed channel separations are $30.7$~km\,s$^{-1}$. The dashed
(red) lines indicate the 1-sigma levels reached for this $19.6$~h
integration.}
\label{fig:sn1997eyhostspec}   
\end{figure}
We observed at IRAM-30m in May 2004 the CO(2-1) line at
146.373~GHz and the CO(3-2) line at 219.55~GHz, relying on the
spectroscopic redshift ($z=0.575$) of the host. At these frequencies,
the telescope's half-power beam widths are respectively 17~\arcsec and
11~\arcsec. We integrated 19.53~h on this source, with typical
system temperatures of 180~K and 260~K (on the T$_A^*$
scale). Wobbler-switching mode was used, with reference positions
offset by 130\arcsec\, in azimuth. For each line, 1-MHz filter bank was
used with a bandwidth of 512~MHz and channel spacing 1~MHz and
the VESPA backend with a bandwidth of 640~MHz and channel spacing
1.25~MHz.

The reduction was performed with the IRAM {\sc GILDAS}
software\footnote{http://www.iram.fr/IRAMFR/GILDAS}. For each line,
the spectra were added and an horizontal baseline was fitted and
subtracted. This simple procedure ensures to avoid possible artifact
due to bad baseline subtraction. The flat resulting spectra, displayed
in Fig. \ref{fig:sn1997eyhostspec}, confirm the good conditions of
observation but also the absence of signal. We then resample the
channels in order to reach 30.7~km\,s$^{-1}$ per channel at
146.373~GHz and 219.55~GHz. Last, we calibrate the spectra using the
standard $S/T_A^*$ factors: 6.7 and 8.7~mJy/K. As displayed in
Fig. \ref{fig:sn1997eyhostspec}, we do not detect any CO lines at the
$3.2$ and $4.7$~mJy (rms) levels respectively.

Given the secure spectroscopic optical redshift and the large
bandwidth at 2~mm, we do not expect a large velocity shift, which
could explain this missing CO emission. Very few galaxies (usually at
$z>3$) present a CO-line width larger than 1000~km\,s$^{-1}$. We would
have detected a non-flat baseline at 2~mm given our reduction
procedure.

Following e.g. \citet{Solomon:2005}, we then compute various upper
limits with different CO-line width $\Delta v$ assumptions as
displayed in Table \ref{tab:upper}: (1) the velocity integrated flux
$S_{\rmn{CO}} \Delta v$ in Jy\,km\,s$^{-1}$; (2) the CO-line luminosity
$L^\prime_{\rmn{CO}}$ expressed in K\,km\,s$^{-1}$\,pc$^2$, computed as:
\begin{equation}
L^\prime_{\rmn{CO}}= 3.25\bmath{\cdot} 10^{7} S_{\rmn{CO}} \Delta v
\frac{D_\rmn{L}^2}{\nu_{\rmn{obs}}^2 (1+z)^3}
\end{equation}
where S$_{\rmn{CO}}$ is the CO flux in Jy, $\Delta v$ is the
(expected) line width in km\,s$^{-1}$, $\nu_{\rmn{rest}}$ is the rest
frequency of the line in GHz, and $D_L$ the luminosity distance in
Mpc; (3) the H$_2$ mass M(${\rmn{H}_2}$)$= \alpha\,
L^\prime_{\rmn{CO}}$ assuming a Milky Way mass-to-CO luminosity ratio
$\alpha=4.6$~M$_\odot$/(K\,km\,s$^{-1}$\,pc$^{2}$); (4) the infrared
luminosity\footnote{Please note that we follow the convention of
\citet{Kewley:2002} for the definition of the infrared luminosity.}
($L_\rmn{IR}$), expressed in L$_\odot$, relying on the relation for
isolated and weakly interacting galaxies from \protect
\citet{Solomon:1988}:
\begin{equation}
L_\rmn{IR} = 3.1\bmath{\cdot} 10^8 {\left(\frac{L^\prime_\rmn{CO}}{10^9}\right)}^{0.74}
\end{equation}
(5) the SFR in M$_\odot$\,yr$^{-1}$ based on \citet{Kewley:2002} empirical
relationship:
\begin{equation}
SFR = 1.7 \bmath{\cdot} 10^{-10} L_{\rmn{IR}}
\end{equation}

We assume that the (expected) CO flux ($S_\rmn{CO}$) is increasing as $\sim
\nu^2$ for the first CO lines, for a given H$_2$ mass, as derived for
starbursts by Combes, Maoli $\&$ Omont \citeyearpar{Combes:1999}. The
ratios $L^\prime_\rmn{CO}(J=2-1)/L^\prime_\rmn{CO}(J=1-0)$ and
$L^\prime_\rmn{CO}(J=3-2)/L^\prime_\rmn{CO}(J=1-0)$ are thus taken to be equal
to 1. This assumes that the lines are thermalized at high temperature
and optically thick. This might not be the case for our host galaxy,
in which case if one relies on studies of nearby galaxies our upper
values should be multiplied by a factor of order 1.1
\citep{Braine:1992} and 1.6 \citep{Devereux:1994}.
\begin{table}
 \centering
  \caption{\protect Upper limits computed for the two studied CO-lines
(IRAM-30m observation). We consider different assumptions for the
expected line widths, and provide $3\sigma$ upper limits on the line
integrated intensity \protect $S_{CO} \Delta v$ and the CO line
luminosity $L^\prime_{CO}$. Assuming a CO-to-H$_2$ ratio appropriate
for the Milky Way, we estimate an upper limit on the molecular mass
$M(\rmn{H}_2)$. Relying on the relation for isolated or
weakly interacting galaxies from \protect \citet{Solomon:1988}, we
provide upper limits on the infrared luminosity $L_\rmn{IR}$. Last,
using the relation of \citet{Kewley:2002}, we derive upper limits on
the SFR.}
\begin{tabular}{lllll} 
\hline
&\multicolumn {4}{l}{Line width $\Delta v$ (km\,s$^{-1}$) $=$}\\ Lines & $100$ &
$250$ & $500$ & $750$\\ \hline 
&\multicolumn{4}{l}{Upper limits on \protect $S_{CO} \Delta v$~\protect(Jy\,km\,s$^{-1}$)}\\ \hline
CO(2-1) & $0.532$ & $0.841$ & $1.19$ & $1.46$\\
CO(3-2) & $0.782$ & $1.24$  & $1.75$ & $2.14$\\
\hline
& \multicolumn {4}{l}{Upper limits on $L^\prime_{CO}$~($10^9$ K\,km\,s$^{-1}$\,pc$^{2}$)}\\ 
 \hline
CO(2-1) & $2.31$ &  $3.65$ & $5.17$ & $6.33$\\ 
CO(3-2) & $1.51$ &  $2.39$ & $3.37$ & $4.13$\\  
\hline
 &\multicolumn {4}{l}{Upper limits on $M(H_2)$
($10^{10}$ M$_\odot$)}\\ 
 \hline
CO(2-1) & $1.06$ &  $1.68$ & $2.38$ & $2.91$\\ 
CO(3-2) & $0.69 $ &  $1.10$ & $1.55$ & $1.9$\\  
\hline
& \multicolumn {4}{l}{Upper limits on $L_{IR}$
($10^{11}$ L$_\odot$)}\\ 
 \hline
CO(2-1) & $1.07$ &  $1.55$ & $2.05$ & $2.41$\\ 
CO(3-2) & $0.76$ &  $1.10$ & $1.45$ & $1.71$\\  
\hline
& \multicolumn {4}{l}{Upper limits on the SFR (M$_\odot$\,yr$^{-1}$)}\\ 
 \hline
CO(2-1) & $32$ &  $47$ & $62$ & $73$\\ 
CO(3-2) & $22.8$ &  $33.1$ & $44$ & $52$\\  
\hline
\end{tabular}
\label{tab:upper}
\end{table}
\begin{figure}
\centering
\includegraphics[width=8cm]{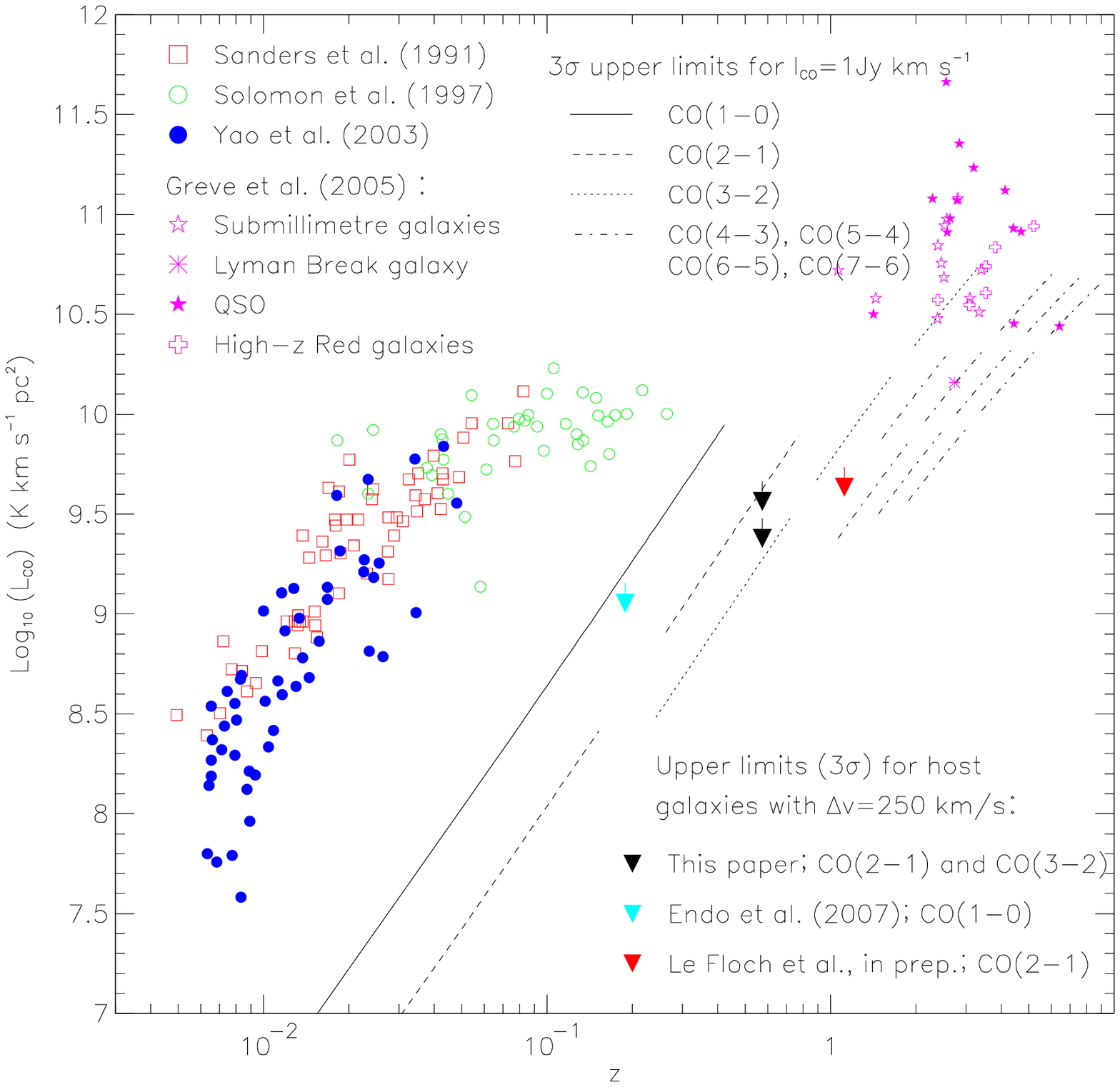}\\
\caption{CO-line luminosities $L^\prime_{CO}$ as a function of
redshift. Upper limit measured here for the studied host is
superimposed on previous measurements from
\protect\citet{Greve:2005}, \citet{Yao:2003}, \citet{Solomon:1997} and
Sanders, Scoville $\&$ Soifer \citeyearpar{Sanders:1991}. We also plot
the upper limits from \protect\citet{Endo:2007} for GRB\,030329 host
and from Le Floc'h (in prep.) for GRB\,000418. None of the CO-line
luminosity has been corrected for gravitational lensing. (This
uncertain correction would lower down some high-z points of this plot
\citep[see e.g.][]{Greve:2005}). The lines indicate the 3$\sigma$
upper limits for $I_{CO}=1$~Jy km s$^{-1}$ and take into account the
atmospheric windows at IRAM.}  
\label{fig:plotgreve}
\end{figure}

Fig. \ref{fig:plotgreve} displays the upper limits derived from our
observations on $L^\prime_\rmn{CO}$, compared to previous detections
of submillimetre galaxies detected in CO
\citep{Greve:2005,Yao:2003,Solomon:1997,Sanders:1991}. We also add for
comparison the upper limits obtained by \citet{Endo:2007} and Le
Floc'h et al. (in prep.) for long GRB hosts. We assumed for all GRB
hosts upper limits based on $\Delta v=250$~km\,s$^{-1}$. This figure
shows that our measurements are competitive with the current state of
the art.

\section[]{Discussion}
\label{label:discussion}
On the basis of the optical spectrum (SFR($[$O\,II$]
\sim 0.2$~M$_\odot$\,yr$^{-1}$), the $B$ luminosity ($0.7$~$L^B_*$)
and the CO lines upper limits, we can securely exclude that this host
galaxy is a strong starburst galaxy. We can derive a star-formation
rate per unit luminosity of 0.3~M$_\odot$\,yr$^{-1}~L^B_*/L_B$. The
SFR and the specific SFR\footnote{If one assumes $M_*/L_B=2$ for a Sc
morphological type \citep{Roberts:1994}, we can estimate a specific SFR of
0.2M$_\odot$\,yr$^{-1}$\,$(10^{10} M_\odot)^{-1}$.} are obviously low
with respect to the global distribution of galaxies \citep[see Figures
17, 18 and 19 in][]{Elbaz:2007} and to the typical submillimetre
galaxies
\citep[with a median SFR of 780\,M$_\odot$\,yr$^{-1}$ in][]{Greve:2005}.

The submillimetre flux is difficult to understand. It has been
obtained by \citet{Farrah:2004} on 2002 December 08, five years after
SN1997ey.  The comparison with templates (see Fig.
\ref{fig:sn1997eyhost}) and the CO lines upper limits suggest that the
SFR could be of order 30~M$_\odot$\,yr$^{-1}$. A factor of order 150
(extinction in B) would then be required to explain the $[$O\,II$]$
luminosity. This submillimeter flux requires a heating source as the
sole interstellar radiation field of the host galaxy cannot heat the
dust up to temperatures close to 20~K \citep{Bethell:2004}.
However, nothing suggests an AGN component in the optical spectrum nor
in the {\it ROSAT} All-Sky Survey (RASS), while the 1.4~GHz upper
limit is compatible with a quiescent or moderate starburst
galaxy. Alternatively, \citet{Clements:2005} discussed that this
strong submillimetre flux could correspond to cirrus, which could
explain the absence of starburst, and derived a (cold-)\,dust mass of
1.3$\bmath{\cdot} 10^9$~M$_\odot$ (T$_{\mathrm{dust}} \sim
20$\,K). This is nevertheless quite far-fetched as this would
correspond to a very large mass of gas: 2$\bmath{\cdot}
10^{11}$~M$_\odot$, assuming a canonical Galactic (cold)dust-to-gas
mass ratios of 1/150, and excluded given our non-detection of the
CO(2-1) line and the excitation temperature of CO(2-1)
(T$_\mathrm{ex}\sim 15$ \,K).

We thus consider two possible explanations: (1) The heating source of
the large gas mass is hidden in the nucleus and escapes
detection. This would require $A_V\sim 4$ to explain the discrepancy
between the [OII] emission line and the submillimetre continuum
flux. (2) This host is member of a cluster and is an anemic or passive
spiral galaxy \citep{Goto:2003}. In this case, one can wonder given
the absence of cross-identification in radio if this submillimetre
flux might not be associated with a background galaxy rather than with
SN1997ey host.  If one considers the 8$\arcsec$\, beam (FWHP) of SCUBA
at 450$\mu$m and integrates up to $z=2$ a Schechter function
\citep[with $\alpha=-1.12$, $\phi_*=0.0128$\,Mpc$^{-3}$,
M$_*=-19.43$,][]{Marzke:1998} with a uniform galaxy distribution, one
expects $1.3$, $4.2$ and $13.6$ galaxies respectively up to $m_B=22$,
$23$ and $24$. In addition, this should be considered as upper limits
as we can observe an over-density in the HST/STIS image. It is thus
possible that the submillimetre source is not associated to SN1997ey
host but to a background object.
\begin{figure}
\centering
\includegraphics[width=8cm]{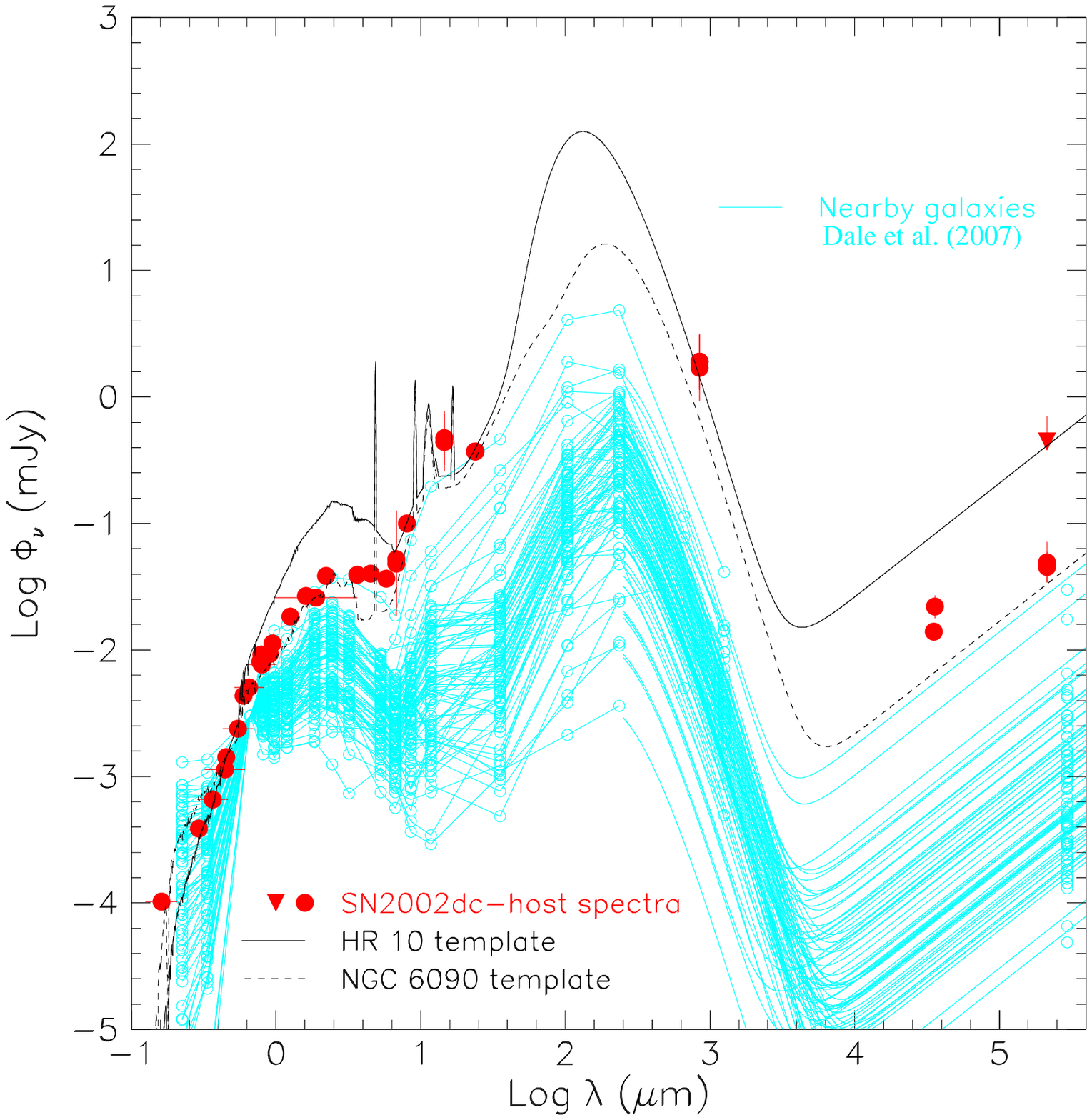}\\
\caption{Same as Fig. \protect\ref{fig:sn1997eyhost} for SN2002dc
host at \protect $z=0.475$ \citep{Cohen:1996}. The $U$, $B$, $V$, $R$, $I$,
$z^\prime$ and $HK^\prime$ data are from
\citet{Capak:2004}, $UBVIJHK$ measurements from Fern{\'a}ndez-Soto,
Lanzetta $\&$ Yahil \protect\citeyearpar{Fernandez-Soto:1999}, $FUV$
(1614$\AA$) measurement from \citet{Teplitz:2006}, the
mid-infrared points from \citet{Pope:2005}, $iz$ and 24~$\mu$m from
\citet{Chary:2005}, $ISO$ measurement from \citet{Goldschmidt:1997}, the
SCUBA point from Wang, Cowie $\&$ Barger \citeyearpar{Wang:2004} and
\citet{Pope:2006} and the 1.4~GHz radio point from
\citet{Richards:2000} and \citet{Wang:2004}.}
\label{fig:sn2002dchost}   
\end{figure}

One can mention the case of the host galaxy of SN2002dc, which has
been detected at $z=0.475$ in the GOODS/ Hubble Deep Field-North field
\citep{Magee:2002,Blakeslee:2003}.  \citet{Pope:2006} has shown that
the submillimetre source HDF850.4 \citep{Wang:2004,Pope:2005} is well
associated with the HDF galaxy 2-264.1
\citep{Williams:1996,Cohen:1996}, which also has near-infrared,
mid-infrared and radio counter-parts. This supernova belongs to the
SNIa sample observed with Spitzer by
\citet{Chary:2005} and is peculiar in the sense that it has the sole
host detected in the submillimetre. This supernova lies at
0.84\arcsec\, from the centre of the spiral galaxy, which corresponds
to a projected distance of 4.9~kpc. Even though this host is in a
starburst phase, it is most probable that the SN exploded in the
outskirts of the starburst zone. Various values of its SFR have been
estimated in the literature and tend to converge towards
25~M$_\odot$\,yr$^{-1}$
\citep{Chary:2005,Pope:2006}, while it is a relatively small mass system with
$L_B\sim 0.3~L_*$. It does not display clear Ca\,II\,H and K
absorption lines, but exhibits strong $[$O\,II$]$ and H$\beta$
emission lines\footnote{See the spectrum in the Hawaii HDF active
catalogue http://www.ifa.hawaii.edu/$\sim$cowie/tts/hdf17.html}. It is
thus a starbursting galaxy, with a star-formation rate per unit
luminosity of 84~M$_\odot$\,yr$^{-1}~L^B_*/L_B$. For this galaxy, all
the observational facts point towards a moderate starburst of
25~M$_\odot$\,yr$^{-1}$ and 0.3~$L^B_*$.

\section{Conclusions}
\label{label:conclusions}
We discussed the properties of SN1997ey host. This SNIa occurred
in a late-type system (0.7~$L_*$). According to the optical data, this
disc galaxy exhibits a residual star-formation activity but no obvious
sign of AGN activity. In parallel, a 6 and 7$\sigma$ submillimetre
flux is detected at 450 and 850~$\mu$m but no heating source
explaining this strong continuum submillimetre flux is detected. We
search for CO lines at $z=0.575$ in this initially promising galaxy
but we have been only able to derive upper limits. We suggest that
either the AGN/starburst activity is hidden by dust in the nucleus or
this host galaxy is anemic or passive and this strong submillimetre
source associated with a background galaxy.

\section*{Acknowledgments}
We thank the referee for her/his detailed comments, which improved the
paper, and F. Daigne for his helpful advices.  We are most grateful to
C. Pennypacker, who encourages us to study SN hosts.  We thank the
Supernova Cosmology Project (SCP) for providing the optical spectrum
displayed in Fig. 1, and specially A. Spadafora who took the time to
get 1997 data from SCP archives, I. Hook and C. Lidman, for providing
insight information about this spectrum.  We thank R. Pain for his
contribution at an earlier stage of this project. We thank L.
Mankiewicz for providing us an early access to the database 'Pi of the
sky'.  We thank the IRAM staff and the 30-m telescope operators for
their assistance with the remote observations performed for this
source.  This research has made use of the NASA/IPAC Extragalactic
Database (NED) which is operated by the Jet Propulsion Laboratory,
California Institute of Technology, under contract with the National
Aeronautics and Space Administration. Some of the data presented in
this paper were obtained from the Multimission Archive at the Space
Telescope Science Institute (MAST). STScI is operated by the
Association of Universities for Research in Astronomy, Inc., under
NASA contract NAS5-26555. Support for MAST for non-{\it{HST}} data is
provided by the NASA Office of Space Science via grant NAG5-7584 and
by other grants and contracts.  This research has made use of the
NASA/ IPAC Infrared Science Archive, which is operated by the Jet
Propulsion Laboratory, California Institute of Technology, under
contract with the National Aeronautics and Space Administration. This
publication makes use of data products from the Two Micron All Sky
Survey, which is a joint project of the University of Massachusetts
and the Infrared Processing and Analysis Centre/California Institute
of Technology, funded by the National Aeronautics and Space
Administration and the National Science Foundation.

\label{lastpage}

\appendix
\section[]{Probability of association with GRB971221}
\label{sec:grb}
\citet{Bosnjak:2006} performed a statistical analysis to determine the
association rate of GRB and SN. They consider both short and long
duration bursts as well as core-collapse and type-Ia
supernovae. Surprisingly, the short burst GRB971221 has thus been
associated with SN1997ey. The burst was detected 1997 December 21,
while the supernova was discovered on December 29 (expected to be the
maximum of the light curve $\pm$ 6 days). As the maximum of SN-Ia
light curves is expected $19.1 \bmath{\cdot} (1+z)$ days after the
explosion \citep{Conley:2006}, the association is plausible.  However,
the error box of 6.3 degree affecting GRB050709 position is so large
that the probability of a chance alignment is huge.  If one integrates
up to $z=2$ a Schechter function \citep[with $\alpha=-1.12$,
$\phi_*=0.0128$\,Mpc$^{-3}$, M$_*=-19.43$,][]{Marzke:1998} with a
uniform galaxy distribution, one expects $7 \bmath{\cdot} 10^6$
(resp. $10^8$) galaxies up to $z=0.5$ (resp. $z=2$) with m$_B$ between
18 and 22 (resp. 24) in the GRB error box. Given the SNIa rate
published by \cite{Sullivan:2006} and assuming an average stellar mass
of the galaxies of $10^{10}$~M$_\odot$, $5.3 \bmath{\cdot} 10^{-4}$
SNIa are expected per year and per galaxy. Given the number of
galaxies present in the GRB error box and assuming a temporal window
of 14 days, one can expect about 140 (resp. 2000) SNIa, while 0.007
short GRB would be expected. We thus consider the association as
highly improbable.

Only a few short GRB \citep[][and references therein]{Levan:2007} were
intensively monitored to exclude definitively some association with
SN.  From a theoretical point of view, short GRB are usually thought
to be generated by the merger of two neutrons stars or one neutron
star with a black hole
\citep[e.g.][]{Paczynski:1991,Narayan:1992}, so no association with SNIa is
expected. However, \cite{King:2001} suggested that the merger of two
white dwarves could lead to the formation of a magnetar, which could
produce short GRB as studied by \cite{Levan:2006}. Interestingly,
coalescing white dwarf binary is one of the channel considered for the
production of SNIa
\citep{Iben:1984,Webbink:1984,King:2001,Ivanova:2006}, compatible with
observations of the population of double white dwarfs in the Galaxy
\citep{Nelemans:2001}. The result of this coalescence depends on the
locus of the carbon ignition, and \cite{Saio:2004} argue that it
probably occurs in the envelope preventing the explosion of a SNIa.







\begin{thebibliography}{99}
\bibitem[Astier et al.(2006)]{Astier:2006} Astier P. et al.\ 
2006, A$\&$A, 447, 31 
\bibitem[Bethell et al.(2004)]{Bethell:2004} Bethell T.~J., 
Zweibel E.~G., Heitsch F., Mathis J.~S.\ 2004, ApJ, 610, 801 
\bibitem[Blakeslee et al.(2003)]{Blakeslee:2003} Blakeslee J.~P. et 
al.\ 2003, ApJ, 589, 693
\bibitem[Bosnjak et al.(2006)]{Bosnjak:2006} Bosnjak Z., Celotti
A., Ghirlanda G., Della Valle M.,  Pian E.\ 2006, A$\&$A, 447, 121 
\bibitem[Braine \& Combes(1992)]{Braine:1992} Braine J., 
Combes F.\ 1992, A$\&$A, 264, 433 
\bibitem[Calzetti et al.(2000)]{Calzetti:2000} Calzetti D., Armus
L., Bohlin R.~C., Kinney A.~L., Koornneef J.,  Storchi-Bergmann T.\ 
2000, ApJ, 533, 682 
\bibitem[Capak et al.(2004)]{Capak:2004} Capak P. et al.\ 2004, 
AJ, 127, 180 
\bibitem[Chary et al.(2005)]{Chary:2005} Chary R., Dickinson
M.~E., Teplitz H.~I., Pope A., Ravindranath S.\ 2005, ApJ, 635, 1022 
\bibitem[Clements et al.(2005)]{Clements:2005} Clements D.~L., 
Farrah D., Rowan-Robinson M., Afonso J., Priddey R., Fox M.\ 2005, 
MNRAS, 363, 229 
\bibitem[Cohen et al.(1996)]{Cohen:1996} Cohen J.~G., Cowie
L.~L., Hogg D.~W., Songaila A., Blandford R., Hu E.~M.,  Shopbell 
P.\ 1996, ApJ, 471, L5 
\bibitem[Combes et al.(1999)]{Combes:1999} Combes F., Maoli R., Omont
A.\ 1999, A$\&$A, 345, 369  
\bibitem[Combes(2004)]{Combes:2004} Combes F.\ 2004, New
Astronomy Review, 48, 583 
\bibitem[Condon et al.(1998)]{Condon:1998} Condon J.~J., Cotton
W.~D., Greisen E.~W., Yin Q.~F., Perley R.~A., Taylor G.~B.,  
Broderick J.~J.\ 1998, AJ, 115, 1693
\bibitem[Conley et al.(2006)]{Conley:2006} Conley A. et al.\ 
2006, AJ, 132, 1707 
\bibitem[Dale et al.(2007)]{Dale:2007} Dale D.~A. et al.\ 2007, 
ApJ, 655, 863 
\bibitem[Delgado et al.(2005)]{Delgado:2005} Delgado R.~M.~G., 
Cervi{\~n}o M., Martins L.~P., Leitherer C.,  Hauschildt P.~H.\ 2005, 
MNRAS, 357, 945
\bibitem[Devereux et al.(1994)]{Devereux:1994} Devereux N., 
Taniguchi Y., Sanders D.~B., Nakai N., Young J.~S.\ 1994, AJ, 107, 
2006 
\bibitem[Dey et al.(1999)]{Dey:1999} Dey A., Graham J.~R., 
Ivison R.~J., Smail I., Wright G.~S., Liu M.~C.\ 1999, ApJ, 519, 
610 
\bibitem[\protect\citeauthoryear{Elbaz et al.}{2007}]{Elbaz:2007} 
Elbaz D., et al., 2007, A\&A, 468, 33 
\bibitem[\protect\citeauthoryear{Endo et al.}{2007}]{Endo:2007}
Endo A., et al., 2007, ApJ, 659, 1431 
\bibitem[Farrah et al.(2002)]{Farrah:2002} Farrah D., Meikle
W.~P.~S., Clements D., Rowan-Robinson M.,  Mattila S.\ 2002, MNRAS, 
336, L17 
\bibitem[Farrah et al.(2004)]{Farrah:2004} Farrah D., Fox M.,
Rowan-Robinson M., Clements D.,  Afonso J.\  2004, ApJ, 603, 489 
\bibitem[Fern{\'a}ndez-Soto et al.(1999)]{Fernandez-Soto:1999} 
Fern{\'a}ndez-Soto A., Lanzetta K.~M.,  Yahil A.\ 1999, ApJ, 513, 34 
\bibitem[\protect\citeauthoryear{Fruchter \& 
Hook}{2002}]{Fruchter:2002} Fruchter A.~S., Hook R.~N. 2002, PASP, 
114, 144 
\bibitem[Goldschmidt et al.(1997)]{Goldschmidt:1997} Goldschmidt P. et 
al.\ 1997, MNRAS, 289, 465 
\bibitem[Greve et al.(2005)]{Greve:2005} Greve T.~R. et al.\ 
2005, MNRAS, 359, 1165 
\bibitem[Hamuy et al.(2003)]{Hamuy:2003} Hamuy M. et al.\
2003,  Nat., 424, 651 
\bibitem[Howell et al.(2005)]{Howell:2005} Howell, D.~A. et al.\ 
2005, ApJ, 634, 1190 
\bibitem[Howell et al.(2007)]{Howell:2007} Howell D.~A.,
Sullivan M., Conley A., Carlberg R.\ 2007, ApJ submitted; preprint
(astro-ph/0701912)
\bibitem[\protect\citeauthoryear{Iben \& 
Tutukov}{1984}]{Iben:1984} Iben I., Jr., Tutukov A.~V., 1984, 
ApJS, 54, 335 
\bibitem[Ivanova et al.(2006)]{Ivanova:2006} Ivanova N., Heinke 
C.~O., Rasio F.~A., Taam R.~E., Belczynski K.,  Fregeau J.\ 2006, 
MNRAS, 372, 1043 
\bibitem[Kewley et al.(2002)]{Kewley:2002} Kewley L.~J., Geller 
M.~J., Jansen R.~A.,  Dopita M.~A.\ 2002, AJ, 124, 3135 
\bibitem[Kewley et al.(2004)]{Kewley:2004} Kewley L.~J., Geller
M.~J., Jansen R.~A.\ 2004, AJ, 127, 2002 
\bibitem[\protect\citeauthoryear{King, Pringle, \& 
Wickramasinghe}{2001}]{King:2001} King A.~R., Pringle J.~E., 
Wickramasinghe D.~T., 2001, MNRAS, 320, L45 
\bibitem[Levan et al.(2006)]{Levan:2006} Levan A.~J., Wynn
G.~A., Chapman R., Davies M.~B., King A.~R., Priddey R.~S.,  Tanvir
N.~R.\ 2006, MNRAS, 368, L1 
\bibitem[Levan et al.(2007)]{Levan:2007} Levan, A.~J., et al.\ 
2007, ArXiv e-prints, 705, arXiv:0705.1705, MNRAS accepted 
\bibitem[McQuade, Calzetti $\&$ Kinney(1995)]{McQuade:1995} McQuade
K., Calzetti D., Kinney A.~L. 
1995, ApJS, 97, 331 
\bibitem[Magee et al.(2002)]{Magee:2002} Magee D. et al.\ 2002, 
IAU Circ. 7908, 1 
\bibitem[Mannucci et al.(2006)]{Mannucci:2006} Mannucci F.,
Della Valle M.,  Panagia N.\ 2006, MNRAS, 370, 773 
\bibitem[Marzke et al.(1998)]{Marzke:1998} Marzke R.~O., da Costa 
L.~N., Pellegrini P.~S., Willmer C.~N.~A., Geller M.~J.\ 1998, ApJ, 
503, 617 
\bibitem[Melchior et al.(2001)]{Melchior:2001} Melchior A.-L., 
Combes F., Guiderdoni B.,  Hatton S.\ 2001, ESA SP-460: The Promise of 
the Herschel Space Observatory, 467; preprint (astro-ph/0102086)
\bibitem[\protect\citeauthoryear{Narayan, Paczynski, \& 
Piran}{1992}]{Narayan:1992} Narayan R., Paczynski B., Piran T.,
1992, ApJ, 395, L83 
\bibitem[\protect\citeauthoryear{Nelemans et 
al.}{2001}]{Nelemans:2001} Nelemans G., Yungelson L.~R., Portegies 
Zwart S.~F., Verbunt F., 2001, A\&A, 365, 491 
\bibitem[Nugent et al.(1998)]{Nugent:1998} Nugent P., Aldering
G., Castro P., Nunes N.,  Quimby R.\ 1998, IAU Circ., 6804, 1 
\bibitem[\protect\citeauthoryear{Osterbrock et 
al.}{1996}]{Osterbrock:1996} Osterbrock D.~E., Fulbright J.~P.,
Martel 
A.~R., Keane M.~J., Trager S.~C., Basri G., 1996, PASP, 108, 277 
\bibitem[\protect\citeauthoryear{Paczynski}{1991}]{Paczynski:1991}
Paczynski B., 1991, AcA, 41, 257 
\bibitem[Pain et al.(2002)]{Pain:2002} Pain R. et al.\ 2002, 
ApJ, 577, 120 
\bibitem[Panagia et al.(2006)]{Panagia:2006} Panagia N., Van
Dyk S.~D., Weiler K.~W., Sramek R.~A., Stockdale C.~J.,  Murata
K.~P.\ 2006, ApJ, 646, 369 
\bibitem[\protect\citeauthoryear{Paturel et 
al.}{1997}]{Paturel:1997} Paturel G. et al.\ 1997,
A\&AS, 124, 109
\bibitem[\protect\citeauthoryear{Peng et al.}{2002}]{Peng:2002} 
Peng C.~Y., Ho L.~C., Impey C.~D., Rix H.-W., 2002, AJ, 124, 266
\bibitem[Perlmutter et al.(1997)]{Perlmutter:1997} Perlmutter S. et 
al.\ 1997, ApJ, 483, 565 
\bibitem[Perlmutter et al.(1999)]{Perlmutter:1999} Perlmutter S. et 
al.\ 1999, ApJ, 517, 565 
\bibitem[Phillips(1993)]{Phillips:1993} Phillips M.~M.\ 1993, ApJ, 
413, L105 
\bibitem[Pope et al.(2005)]{Pope:2005} Pope A., Borys C., 
Scott D., Conselice C., Dickinson M.,  Mobasher B.\ 2005, MNRAS, 
358, 149
\bibitem[Pope et al.(2006)]{Pope:2006} Pope A. et al.\ 2006, 
MNRAS, 370, 1185 
\bibitem[Richards(2000)]{Richards:2000} Richards E.~A.\ 2000, ApJ, 
533, 611 
\bibitem[Riess et al.(1996)]{Riess:1996} Riess A.~G., Press 
W.~H., Kirshner R.~P.\ 1996, ApJ, 473, 88 
\bibitem[Riess et al.(1998)]{Riess:1998} Riess A.~G. et al.\ 
1998, AJ, 116, 1009 
\bibitem[Roberts \& Haynes(1994)]{Roberts:1994} Roberts, M.~S., \& 
Haynes, M.~P.\ 1994, ARAA, 32, 115 
\bibitem[Rose(1985)]{Rose:1985} Rose J.~A.\ 1985, AJ, 90, 1927 
\bibitem[Sadat et al.(1998)]{Sadat:1998} Sadat R., Blanchard A., 
Guiderdoni B.,  Silk J.\ 1998, A$\&$A, 331, L69 
\bibitem[\protect\citeauthoryear{Saio \& 
Nomoto}{2004}]{Saio:2004} Saio H., Nomoto K., 2004, ApJ, 615, 444 
\bibitem[Sanders et al.(1991)]{Sanders:1991} Sanders D.~B., 
Scoville N.~Z.,  Soifer B.~T.\ 1991, ApJ, 370, 158 
\bibitem[\protect\citeauthoryear{Schwope et 
al.}{2000}]{Schwope:2000} Schwope A., et al., 2000, AN, 321, 1 
\bibitem[Skrutskie et al.(2006)]{Skrutskie:2006} Skrutskie M.~F. et 
al.\ 2006, AJ, 131, 1163
\bibitem[Solomon \& Sage(1988)]{Solomon:1988} Solomon P.~M., 
Sage L.~J.\ 1988, ApJ, 334, 613 
\bibitem[Solomon \& Vanden Bout(2005)]{Solomon:2005} Solomon P.~M., 
 Vanden Bout P.~A.\ 2005, ARAA, 43, 677 
\bibitem[Solomon et al.(1997)]{Solomon:1997} Solomon P.~M., Downes
D., Radford S.~J.~E., Barrett J.~W.\ 1997, ApJ, 478, 144 
\bibitem[\protect\citeauthoryear{Goto et al.}{2003}]{Goto:2003} 
Goto T. et al.\ 2003, PASJ, 55, 757 
\bibitem[\protect\citeauthoryear{Spergel et al.}{2007}]{Spergel:2007}
Spergel D.~N., et al., 2007, ApJS, 170, 377 
\bibitem[Sullivan et al.(2003)]{Sullivan:2003} Sullivan M. et al.\ 
2003, MNRAS, 340, 1057 
\bibitem[Sullivan et al.(2006)]{Sullivan:2006} Sullivan M. et
al.\ 2006, ApJ, 648, 868 
\bibitem[Teplitz et al.(2006)]{Teplitz:2006} Teplitz H.~I. et al.\ 
2006, AJ, 132, 853 
\bibitem[Wang et al.(2004)]{Wang:2004} Wang W.-H., Cowie L.~L., 
 Barger A.~J.\ 2004, ApJ, 613, 655 
\bibitem[Williams et al.(1996)]{Williams:1996} Williams R.~E. et 
al.\ 1996, AJ, 112, 1335 
\bibitem[Yao et al.(2003)]{Yao:2003} Yao L., Seaquist E.~R., 
Kuno N., Dunne L.\ 2003, ApJ, 588, 771 
\bibitem[\protect\citeauthoryear{Webbink}{1984}]{Webbink:1984} 
Webbink R.~F., 1984, ApJ, 277, 355 
\end{thebibliography}
\end{document}